\documentclass[%
 reprint,
 amsmath,amssymb,
 aps,
]{revtex4-1}

\usepackage{graphicx}
\usepackage{dcolumn}
\usepackage{bm}

\begin{document}

\preprint{APS/123-QED}

\title{All-fiber all-normal-dispersion femtosecond laser with nonlinear multimodal interference based saturable absorber}

\author{U\u{g}ur Te\u{g}in}
 \email{ugurtegin@unam.bilkent.edu.tr}
\author{B\"{u}lend Orta\c{c}}%
 \email{ortac@unam.bilkent.edu.tr}
\affiliation{%
UNAM — National Nanotechnology Research Center and\\ Institute of
Materials Science and Nanotechnology, Bilkent University
}%

\begin{abstract}
In this letter, we demonstrate the first all-fiber all-normal dispersion ytterbium-doped oscillator with nonlinear multimodal interference based saturable absorber capable to generate ultrashort dissipative soliton pulses. Additional to functioning as a saturable absorber, the use of multimode fiber segments between single mode fibers also ensures the bandpass filtering via multimode interference reimaging necessary to obtain dissipative soliton mode-locking. The oscillator generates dissipative soliton pulses at 1030 nm with 5.8 mW average power, 5 ps duration and 44.25 MHz repetition rate. Pulses are dechirped to 276 fs via an external grating compressor. All-fiber cavity design ensures high stability and $\sim$70 dB sideband suppression is measured in RF spectrum. Numerical simulations are performed to investigate cavity dynamics and obtained results well-matched with experimental observations. The proposed cavity presents an alternative approach to achieve all-fiber dissipative soliton mode-locking with a simple and low-cost design.
\end{abstract}

\maketitle
Fiber-based laser systems are important for material processing, medical applications and optical metrology \cite{fermann2013ultrafast}. Ytterbium based fiber lasers are generally preferred in the photonic systems due to high and broadband gain of the ytterbium active ions. Various mode-locking dynamics are proposed in the literature to obtain ultrashort pulses from ytterbium based fiber cavities with soliton \cite{isomaki2006femtosecond}, dispersion managed \cite{ortacc200390} and self-similar pulses \cite{dudley2001generation,ilday2004self}. However, a certain degree of dispersion mapping inside the laser cavity is necessary to achieve stable mode-locking operation in these type of fiber lasers. At 1$\mu$m wavelength range negative dispersion can be implemented by bulk grating or photonic crystal fibers which results in increased complexity and undermines benefits of fiber lasers such as compactness and stability. Later, a stable passively mode-locked all-normal dispersion fiber laser is demonstrated and pulse generation is attributed to the strong spectral filtering of chirped pulses \cite{chong2006all}. Over the last decade, all-normal dispersion fiber lasers are studied extensively by exploiting dissipative soliton pulse dynamics \cite{grelu2012dissipative}. In the literature, the power and energy scalability of the dissipative soliton pulses is demonstrated with very-large-mode-area fibers \cite{ortacc2009approaching,baumgartl201266}. On the other hand, all-fiber dissipative soliton lasers are a subject of high interest owing to their compact and misalignment free designs. In the literature, researchers demonstrate different all-fiber oscillators by proposing several methods to obtain in-line all-fiber bandpass filtering necessary to achieve dissipative soliton pulses at 1$\mu$m wavelength range \cite{kieu2008all,schultz2008all,ozgoren2010all,zhang2012tunable}. Nonlinear polarization evolution (NPE) and material based saturable absorbers are generally preferred to start mode-locked operation in these studies.

In the last few years, graded-index multimode fibers become subject to the extensive study of nonlinear optics due to their unique properties. In the literature, researchers reported new nonlinear dynamics such as spatiotemporal instability \cite{krupa2016observation,wright2016self,teugin2017spatiotemporal}, supercontinuum generation \cite{lopez2016visible,krupa2016spatiotemporal,teugin2017high}, self-beam cleaning \cite{lopez2016visible,krupa2016spatial,liu2016kerr}, multimode solitons \cite{renninger2013optical} and their dispersive waves \cite{wright2015controllable}. In addition to the aforementioned studies, saturable absorber behavior of a short graded-index multimode fiber segment is theoretically proposed by Nazemosadat et al. \cite{nazemosadat2013nonlinear}. In the proposed method, in between single mode fiber segments, a graded-index multimode fiber can be used to obtain nonlinear multimodal interference (NL-MMI) to achieve saturable absorber attribute. The results of Nazemosadat et al. indicate a saturable absorber can be obtained with more than 90\% modulation depths via proposed structure. On the other hand, an all-fiber bandpass filter is later demonstrated with similar device structure both numerically and experimentally \cite{mohammed2006all}. The proposed filter is based on multimode interference reimaging phenomenon. Later, Mafi et al. presented a low-loss wavelength dependent coupler structure by implementing a graded-index multimode fiber segment to this filter structure \cite{mafi2011low}.Very recently, the use of a short graded-index multimode fiber segment as a saturable absorber is experimentally verified in a thulium based all-fiber soliton oscillator with a slightly modified structure \cite{li2017mode}. In between the single mode and graded-index multimode fibers, Li et al. placed a step-index multimode fiber to ensure excitation of the high order modes in the graded-index multimode fiber and achieved saturable absorber behavior. They reported a soliton all-fiber oscillator capable to generate 1.4 ps pulses at 1888 nm with 19.82 MHz repetition rate and 0.6 mW output average powers. Additionally, they systematically studied the effect of step-index and graded-index multimode fiber lengths in the proposed saturable absorber structure to the mode-locking threshold and output power of the laser. The proposed NL-MMI based saturable absorber configuration also supports bandpass filter structure thus it is very promising for all-normal dispersion lasers to obtain dissipative soliton pulses with simple and all-fiber cavity design. However, to the best of our knowledge, there are no reports on employing the NL-MMI based saturable absorber and bandpass filter configuration of a multimode fiber segment in all-normal-dispersion mode-locked fiber laser to realize a dissipative-soliton operation.

\begin{figure}[t!]
\centering
\includegraphics[width=\linewidth]{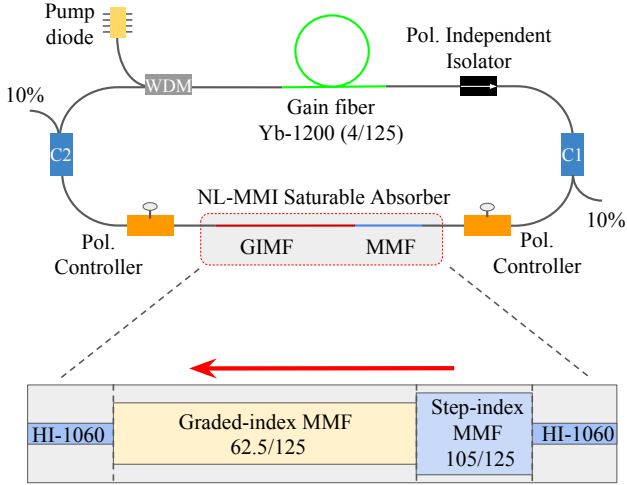}
\caption{Schematic of the all-fiber Yb-doped laser with NL-MMI based saturable absorber: WDM, wavelength division multiplexer; GIMF, graded-index multimode fiber; MMF, multimode fiber.}
\label{Setup}
\end{figure}

Here we present the first all-fiber integrated all-normal dispersion Yb-doped oscillator with NL-MMI based saturable absorber capable to generate ultrashort pulses both numerically and experimentally. With this simple cavity design, a compact stable and low-cost dissipative soliton fiber laser is presented. Numerical simulations investigate the possibility of dissipative soliton mode-locking and lead to experimental studies by revealing the internal cavity dynamics. Experimentally demonstrated oscillator is self-starting and generates dissipative solitons at 1030 nm with 5.8 mW average power and 44.25 MHz repetition rate. Chirped dissipative soliton pulses are compressed externally to 276 fs. 

The oscillator scheme is presented in Fig. \ref{Setup}. The length of the multimode fibers used in saturable absorber segment is important for mode-locking threshold and bandwidth of the bandpass filter. As illustrated in Fig. \ref{Setup}, we prefer to use the step-index multimode fiber (105/125) to graded-index multimode fiber (62.5/125) configuration as described by Li et al. \citep{li2017mode} for adequate excitation of graded-index multimode fiber. To obtain $\sim$ 15 nm bandwidth for multimode interference based bandpass filtering, the length of the graded-index multimode fiber is determined as 14 cm  \citep{mohammed2006all,mafi2011low}. With 0.5 cm step-index multimode fiber segment, the total length of the NL-MMI based saturable absorber segment is defined as 14.5 cm. The transmission behavior of NL-MMI based saturable absorber is presented as sinusoidal by Nazemosadat et al. \citep{nazemosadat2013nonlinear}. According to Nazemosadat et al. the modulation depth and minimum transmission of saturable absorber segment are defined by the length of graded-index multimode fiber and energy splitting between the modes thus for a constant fiber length, one can achieve desired modulation depth. 

\begin{figure}[b!]
\centering
\includegraphics[width=\linewidth]{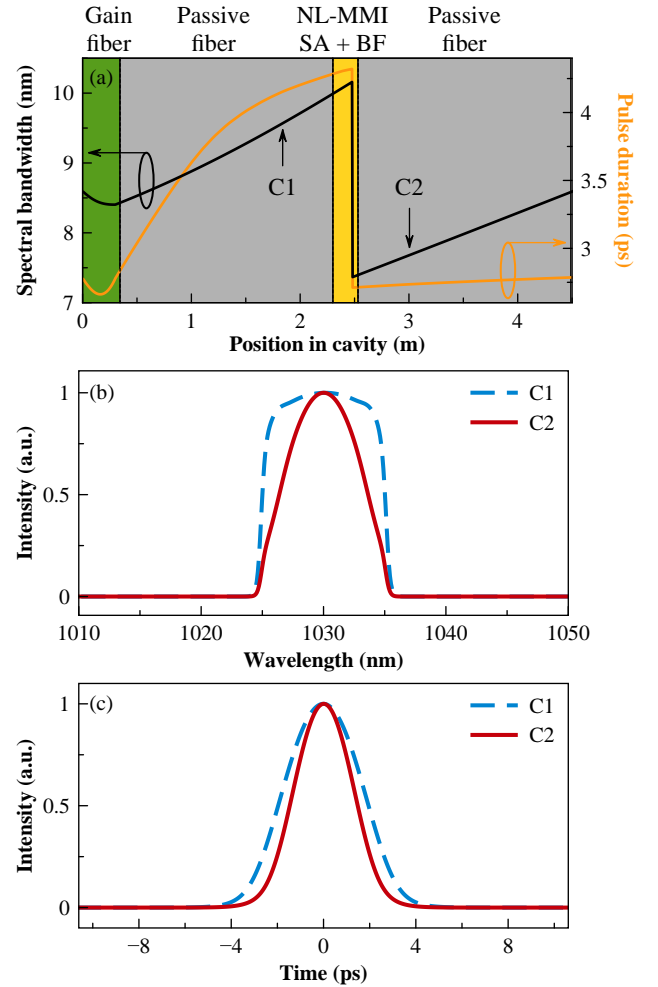}
\caption{(a) Simulated pulse duration and spectral bandwidth variation over the cavity: SA, saturable absorber; BF, bandpass filter. (b) Simulated laser spectra obtained from the defined output couplers. C1 (C2) is defined after (before) the gain fiber segment. (c) Simulated temporal profile obtained at the output couplers C1 and C2.}
\label{Sim3}
\end{figure}

We first performed numerical simulations based on the model described in \citep{agrawal2007nonlinear,agrawal2001applications,csenel201333} to investigate the possibility of dissipative soliton mode-locking with the proposed cavity (Fig. \ref{Setup}). Based on Nazemosadat et al.'s study, we modelled the saturable absorber segment by a transfer function of the form $T = q_{0}+q_{1}sin^{2}(I/I_{sat})$ where $q_{0} = 0.15$ is the minimum transmission, $q_{1} = 0.9$ is the modulation depth, $I$ is instantaneous pulse power and $I_{sat}$ is the saturation power \citep{nazemosadat2013nonlinear}. In the NL-MMI saturable absorber, after the graded-index multimode fiber light propagates into a single mode fiber therefore the energy in the higher order modes act as an additional intracavity loss. Thus in the simulations, we only consider pulse propagation in the fundamental mode and introduce an additional loss term at the saturable absorber segment. This loss term mimics the energy content of the high order modes which cannot propagate in the following single mode fiber segment. By comparing the fiber core sizes of the graded-index multimode and single mode fibers, we defined this loss term as $0.95$ in our simulations. With 14 cm graded-index multimode fiber segment of the cavity, multimode interference bandpass filter is defined as a Gaussian filter with 15 nm bandwidth according to calculations presented by Mafi et al. \citep{mafi2011low}. In simulations, the saturable absorber transfer function, bandpass filtering and the loss term are introduced at the end of the graded-index multimode fiber. The gain is defined as Lorentzian with 40 nm bandwidth and 30 dB small-signal gain. Raman scattering and shock terms are also incorporated in our numerical studies. 

\begin{figure}[t!]
\centering
\includegraphics[width=\linewidth]{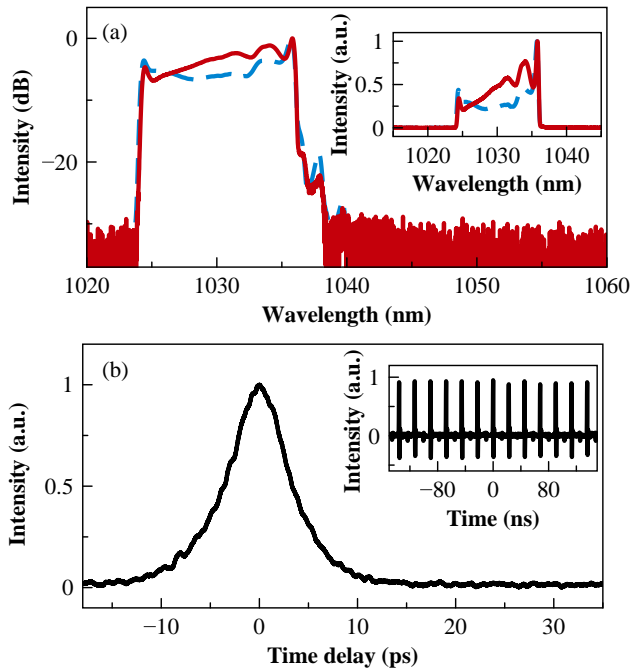}
\caption{Spectra of dissipative soliton pulse from the output couplers on logarithmic (a) and linear scale (inset). Measurements from C1 and C2 couplers are indicated as blue and red, respectively. (b) Autocorrelations trace of the chirped pulses obtained directly from output coupler C1. Inset: Single-pulse train.}
\label{OSA}
\end{figure}

The numerical simulations are performed with fourth-order Runge–Kutta in the interaction picture method \citep{hult2007fourth,csenel201333}. The initial field is defined as a quantum noise. A stable mode-lock regime is found and the results are demonstrated in the Fig. \ref{Sim3}. At the first output port after the gain fiber (C1) dissipative soliton pulses with 1.3 nJ intracavity pulse energy are obtained. Intracavity evolution of the pulse width in the spectral and temporal domain is presented in Fig. \ref{Sim3}(a). We notice that gain filtering shortens the spectral width in the first part of the active fiber, whereas the spectrum is broadened by Kerr nonlinearity in following passive fiber segment. The required pulse shaping based on chirping and spectral filtering is acquired by multimode interference based bandpass filter segment. The relatively high intracavity loss due to multimode to single mode fiber propagation and strong amplification support the self-consistent dissipative soliton evolution. For output couplers before and after the gain fiber, the spectra have 10 nm and 7.3 nm bandwidths respectively [Fig. \ref{Sim3}(b)]. As presented in Fig. \ref{Sim3}(c), the output pulse duration of 3.9 ps and 2.9 ps obtained respectively for these couplers. The same numerical solution is reached from different initial quantum noise fields.

\begin{figure}[t!]
\centering
\includegraphics[width=\linewidth]{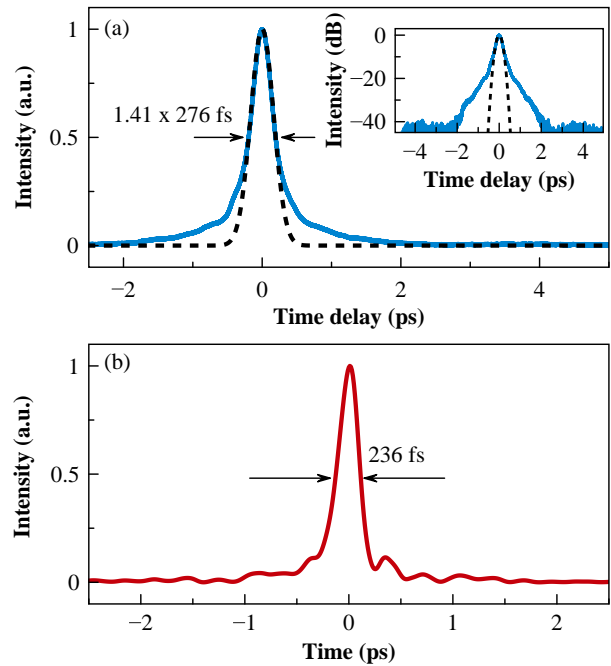}
\caption{(a) Autocorrelation trace of the dechirped pulses obtained from output coupler C1 (solid) and theoretical fit with Gaussian pulse shape (dashed). Inset: In logarithmic scale. (b) PICASO retrieved dechirped pulse shape.}
\label{Pulse}
\end{figure}

The experiments are then performed with the cavity design determined by the numerical simulations. The gain section of the oscillator is 30 cm highly doped ytterbium fiber pumped by 976 nm fiber-coupled diode laser. An inline isolator is employed to determine pulse propagation direction inside the ring cavity. To eliminate the possibility of NPE mode-locking we intentionally prefer to use a polarization independent isolator. As defined by the simulations, 10\% output couplers are placed before and after the NL-MMI based saturable absorber segment to investigate intracavity pulse dynamics. NL-MMI saturable absorber segment is developed by 0.5 cm step-index multimode fiber with 105 $\mu$m core and 125 $\mu$m cladding diameter (Nufern MM-S105/125) and 14 cm graded-index multimode fiber with 62.5 $\mu$m core and 125 $\mu$m cladding diameter (Thorlabs GIF625). Fiber polarization controllers are used for ensuring optimum intra-cavity birefringence to achieve mode-locking operation. Self-starting fundamental mode locking of single pulse train with a repetition rate of $\sim$44 MHz observed around 400 mW pump power. Measured optical spectra from the output couplers are presented in Fig. \ref{OSA}(a) and the central wavelength of the pulses are recorded at 1030 nm. Spectral bandwidths are recorded as 11.8 nm and 11.4 nm at -7 dB from the couplers C1 and C2, respectively. Based on this observation, the pulse breathing ratio in one round trip is close to unity which is typical behavior of dissipative solitons in all-normal dispersion fiber lasers. The output power of the laser is measured as 5.8 mW and 0.15 mW at the couplers C1 and C2, respectively. As we expect from the numerical simulations, the intracavity pulse energy of 1.3 nJ obtained experimentally before the output coupler (C1). The temporal characterizations are performed for the pulses obtained from C1 output coupler. The laser produces chirped pulses with $\sim$ 5 ps duration as presented in the autocorrelation trace in Fig. \ref{OSA}(b). These chirped pulses are compressed externally using a grating compressor with a 300 line/mm diffraction grating pair. The minimum pulse duration of 276 fs is obtained with Gaussian deconvolution factor and demonstrated in Fig. \ref{Pulse}(a). To retrieve the temporal profile of the compressed pulses from the autocorrelation and spectrum data, we employ PICASO algorithm \cite{nicholson1999full}. The PICASO-retrieved shape of the pulse is presented in the Fig. \ref{Pulse}(b). The algorithm indicates retrieved pulse duration is 236 fs.

\begin{figure}[t!]
\centering
\includegraphics[width=\linewidth]{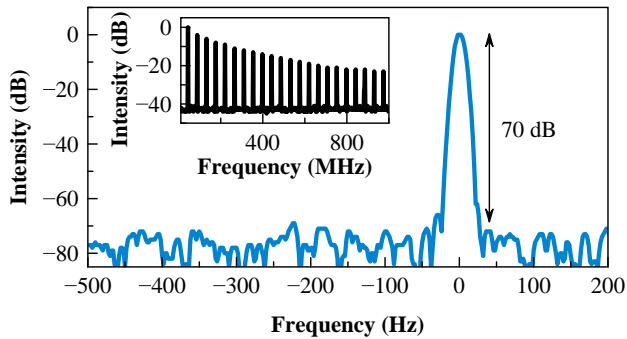}
\caption{Measured RF spectrum with 1 kHz span and 10 Hz resolution bandwidth, with central frequency shifted to zero for clarity. Inset, RF spectrum with 1 GHz span and 3 kHz resolution bandwidth.}
\label{RF}
\end{figure}

We characterized the oscillator in the frequency domain as well. The fundamental repetition rate of the laser is verified with a with radio frequency (RF) analyzer as 44.25 MHz. With 1 kHz span and 10 Hz resolution bandwidth, we measured RF spectrum and observe sideband suppression ratio around 70 dB (Fig. \ref{RF}). The fiber laser has outstanding stability both in the short and long term. The laser continues mode-locking operation uninterrupted for days and there is no sign of degradation. The measured RF spectrum with 1 GHz span and 3 kHz resolution bandwidth indicates no periodic envelope or fluctuation observed in long-span.

In conclusion, we numerically and experimentally demonstrate an all-fiber all-normal dispersion ytterbium-doped oscillator with nonlinear multimodal interference based saturable absorber capable to generate ultrashort dissipative soliton pulses, for the first time in the literature. Via multimode interference reimaging, the preferred saturable absorber structure also ensures the bandpass filtering necessary to obtain dissipative soliton pulses. Thus the proposed cavity presents an alternative approach to achieve all-fiber dissipative soliton mode-locking with the simple and low-cost design. The oscillator generates dissipative soliton pulses at 1030 nm with 5.8 mW average power, 5 ps duration and 44.25 MHz repetition rate. Pulses are dechirped to 276 fs via an external grating compressor. All-fiber cavity design ensures high stability. The low-noise operation of the all-fiber cavity is confirmed and $\sim$70 dB sideband suppression is measured in RF spectrum. We believe the reported all-fiber oscillator configuration is an attractive solution for generation ultrashort dissipative soliton pulses and can find application in supercontinuum generation and optical metrology.

\bibliography{sample}

\end{document}